\documentclass[a4paper,11pt]{article}
\usepackage{pos}

\title{Measurements of $|V_{cb}|$ and $|V_{ub}|$ from Belle and Belle~II}
\author*[a]{Lu Cao}

\affiliation{Deutsches Elektronen-Synchrotron (DESY),\\
 Notkestraße 85, Hamburg, Germany}
  
\note{On behalf of the Belle and Belle~II collaborations.}

\emailAdd{lu.cao@desy.de}

\abstract{This proceeding summarizes recent measurements of the Cabibbo-Kobayashi-Maskawa (CKM) matrix elements $|V_{cb}|$ and $|V_{ub}|$ at the Belle and Belle~II experiments. It provides insights derived from both exclusive and inclusive determinations. Preliminary results from Belle~II are discussed, focusing on $B^0\to D^{*-}\ell^+ \nu$ and $B^{0} \to \pi^{-} \ell^{+} \nu$ utilizing $189 \, \mathrm{fb}^{-1}$ data collected at the $\Upsilon(4S)$ resonance. This reveals the derived branching fractions and $|V_{xb}|$. The shape measurement of $B \to D^{*} \ell \nu$ at Belle extracted $|V_{cb}|$ employing new lattice-QCD calculations as input and provided additional tests on the universality of light lepton flavors. In the realm of inclusive decay, the measurement for $|V_{ub}|$ and the differential spectra of $B \to X_{u} \ell \nu$ decays have been performed. This presentation also encapsulates the new analyses of $|V_{ub}|^{\mathrm{excl}}/|V_{ub}|^{\mathrm{incl}}$ and $|V_{ub}|^{\mathrm{incl}}/|V_{cb}|^{\mathrm{incl}}$ ratios.}

\FullConference{20th International Conference on B-Physics at Frontier Machines (Beauty2023)\\
 3-7 July, 2023\\
Clermont-Ferrand, France\\}

\begin{document}
\maketitle

% One of the crucial tests of the Standard Model of particle physics (SM) is a precise determination of the magnitude of the Cabibbo-Kobayashi-Maskawa (CKM) matrix elements describing the quark mixing and accounts for $CP$-violation in the quark sector. However, the disagreement between the exclusive and inclusive determinations of $|V_{xb}|$ is about three standard deviations \cite{Amhis:2019ckw}. The pursuit of an in-depth investigation into precision measurements through the utilization of new experimental data and advanced techniques remains an important topic in this field.

\section{Exclusive $|V_{cb}|$}

\begin{figure}[h!]
    \centering
    \includegraphics[width=0.4\columnwidth]{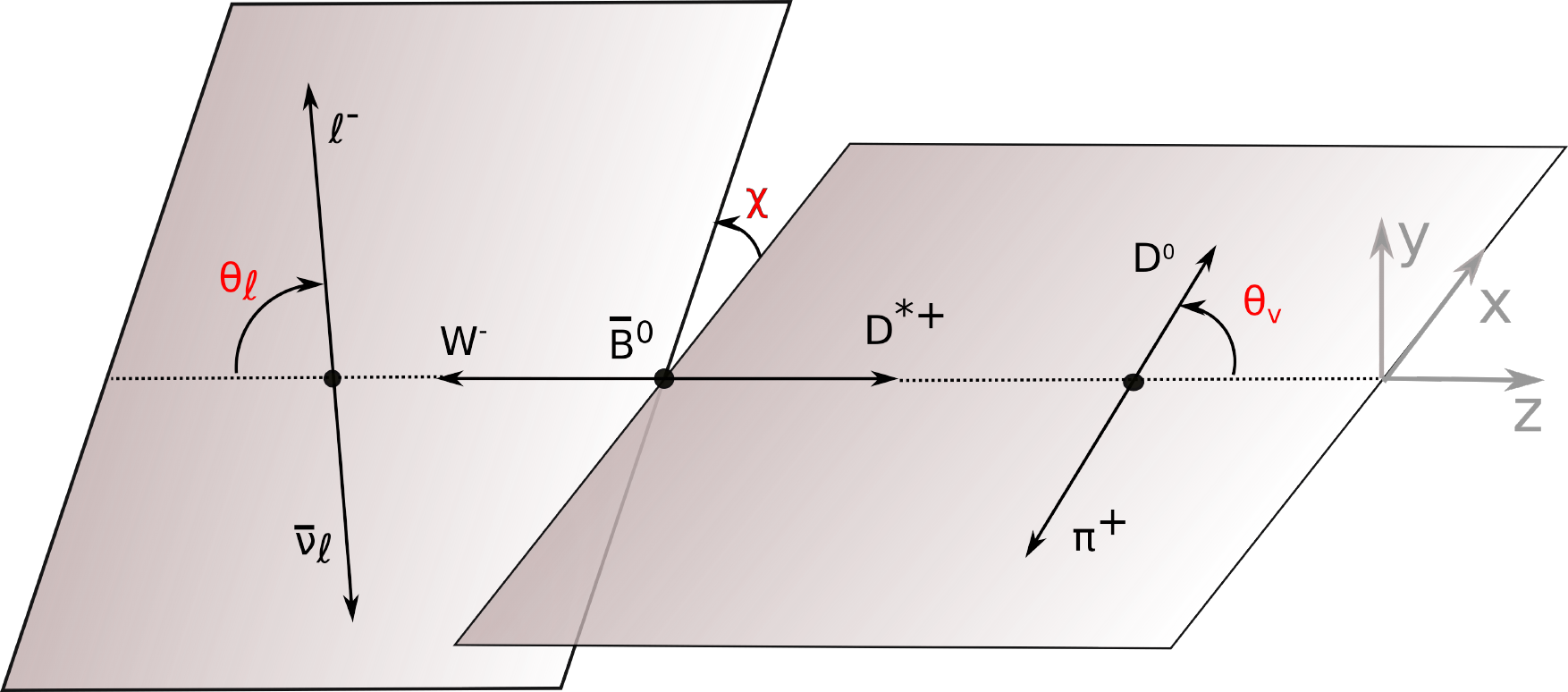}
    \caption{Sketch of the helicity angles $\theta_\ell$, $\theta_V$, and $\chi$ that characterize the $B \to D^{*} \ell \nu$ decay.}
    \label{fig:helicity}
\end{figure}

%% Chaoyi's
With the collected $189 \, \mathrm{fb}^{-1}$ of collision data at the Belle~II experiment, $B^0\to D^{*-} \ell^{+} \nu$ ($\ell = e, \mu$) decays are reconstructed to determine the distributions of hadronic recoil parameter $w = (m^{2}_{B} + m^{2}_{D^{*}} - q^{2})/(2 m_{B}m_{D^{*}})$ and three decay angles $\theta_\ell$, $\theta_V$, and $\chi$ (see Fig.~\ref{fig:helicity}). The partial decay rates are measured as functions of the four kinematic variables separately for electron and muon final states. We obtain $|V_{cb}|$ using the Boyd-Grinstein-Lebed~\cite{Boyd:1995sq,Boyd:1997kz} (BGL) and Caprini-Lellouch-Neubert~\cite{Caprini:1997mu} (CLN) parametrizations, and find $|V_{cb}|_\mathrm{BGL}=(40.57\pm 0.31 \pm 0.95\pm 0.58 )\times 10^{-3}$ and $|V_{cb}|_\mathrm{CLN}=(40.13 \pm 0.27 \pm 0.93\pm 0.58 )\times 10^{-3}$ with the uncertainties denoting statistical, systematic components, and source from the lattice QCD input, respectively. We employ a nested hypothesis test, based on the proposal in Ref.~\cite{Bernlochner:2019ldg}, to determine the order at which to truncate the expansion of BGL form factors. The $|V_{cb}|$ result is in good agreement with the world average of the exclusive approach and the inclusive determination of Refs.~\cite{Bordone:2021oof,Bernlochner:2022ucr}. We assess the effect of incorporating nonzero recoil information from Ref.~\cite{FermilabLattice:2021cdg} into two scenarios using the same order of BGL expansion, observing a slight decrease in the calculated value of $|V_{cb}|$ when considering $h_{A_1}$ and tension with FNAL/MILC lattice predictions when including all form factors. The measured branching ratio is $\mathcal{B}(B^0\to D^{*-} \ell^{+} \nu)=(4.922 \pm 0.023 \pm 0.220)\%$, and the ratio for electron and muon mode is $0.998 \pm 0.009 \pm 0.020$. Additionally, we observe the forward-backward angular asymmetry and $D^{*+}$ longitudinal polarization fractions, all consistent with Standard Model lepton-flavor universality. The preliminary results of this analysis have been recently published in Ref.~\cite{chaoyi}.

%% Markus
In a recent Belle measurement \cite{Belle:2023bwv}, the exclusive $|V_{cb}|$ was determined by analyzing the differential shapes of $B\to D^* \ell \bar{\nu}\ell$ decays ($B = B^-, \bar{B}^0 $) with hadronic tag-side reconstruction, utilizing the full Belle data set of $711,\mathrm{fb}^{-1}$ integrated luminosity. The signal yields are extracted in 160 differential bins, characterizing the 1D projections of the hadronic recoil parameter $w$ and the three helicity angles. With an external input for the absolute branching fractions, the decay form factor parameters are obtained by fitting the kinematic differential shapes of this decay, and the number of floating BGL parameters is determined using a nested hypothesis test~\cite{Bernlochner:2019ldg}. The resulting $|V_{cb}|$ values are $|V_{cb}|_\mathrm{CLN} = (40.1\pm0.9)\times 10^{-3}$ and $|V_{cb}|_\mathrm{BGL} = (40.6\pm 0.9)\times 10^{-3}$ with the zero-recoil lattice QCD point $\mathcal{F}(1) = 0.906 \pm 0.013$. This result agrees with the Belle~II measurement~\cite{chaoyi} and also the recent inclusive determinations~\cite{Bordone:2021oof,Bernlochner:2022ucr}. A study of the impact of preliminary beyond zero-recoil lattice QCD calculations~\cite{FermilabLattice:2021cdg} on the $|V_{cb}|$ determinations is performed. Furthermore, we presented the lepton flavor universality ratio $R_{e\mu} = \mathcal{B}(B \to D^* e \bar{\nu}_e) / \mathcal{B}(B \to D^* \mu \bar{\nu}_\mu) = 0.990 \pm 0.021 \pm 0.023$, the electron and muon forward-backward asymmetry and their difference $\Delta A_{FB}=0.022\pm0.026\pm 0.007$, and the electron and muon $D^*$ longitudinal polarization fraction and their difference $\Delta F_L^{D^*} = 0.034 \pm 0.024 \pm 0.007$, all consistent with SM predictions.

\section{Exclusive $|V_{ub}|$}

%Svenja
The charmless semileptonic decay $B^0\to\pi^- \ell^+ \nu$ and exclusive $|V_{ub}|$ are measured by Belle~II~\cite{Belle-II:2022imn} utilizing the primary data set of 198.0 million pairs of $B\bar{B}$ mesons recorded at the Belle~II detector~\cite{Abe:2010sj} at a center-of-mass (CM) energy of $\sqrt{s} = $ 10.58~GeV. The $B^0\to\pi^- \ell^+ \nu$ decay is reconstructed without identifying the partner $B$ meson. The reconstructed events are separated into six distinct intervals (bins) of squared momentum transfer $q^{2}$. The signal yields are extracted through an extended likelihood fit applied to a binned two-dimensional distribution of the energy difference $\Delta E = E^{*}_{B} - E^{*}_{\mathrm{beam}} = E^{*}_{B} - \frac{\sqrt{s}}{2 }$ and the beam-constrained mass $M_{bc}=\sqrt{E^{*2}_{\mathrm{beam}} -|\vec{p}^{\,*}_{B}|^{2}} =\sqrt{\left(\frac{\sqrt{s}}{2 }\right)^{2} -|\vec{p}^{\,*}_{B}|^{2}}$ in each $q^{2}$ bin. Here, $E^{*}_{\mathrm{beam}}$, $E^{*}_{B}$ and $\vec{p}^{\,*}_{B}$ are the single-beam energy, reconstructed $B$ energy, and reconstructed $B$ momentum all determined in the $\Upsilon(4S)$ rest frame, respectively. The partial branching fractions are measured independently for $B^0\to\pi^- e^+ \nu$ and $B^0\to\pi^- \mu^+ \nu$ as functions of $q^{2}$. The total branching fraction is found to be $(1.426 \pm 0.056 \pm 0.125) \times 10^{-4}$ for $B^0\to\pi^- \ell^+ \nu$ decays. The decay form factors are based on Bourrely-Caprini-Lellouch (BCL) parameterization~\cite{Bourrely:2008za}. By fitting the measured differential $q^{2}$ spectra, and incorporating constraints from lattice QCD calculations~\cite{FermilabLattice:2015mwy}, the value of $|V_{ub}|$ is obtained as $(3.55 \pm 0.12 \pm 0.13 \pm 0.17) \times 10^{-3}$ with statistical, systematic, and theoretical uncertainties separated. The dominant sources of systematic uncertainty arose from the modeling of background contributions from continuum and $B \to \rho \ell \nu$ events.

\section{Inclusive $|V_{ub}|$}

The recent Belle measurement~\cite{cao:2021prd} determined the partial branching fraction of inclusive semileptonic $B\to X_{u} \ell \nu$ decays with hadronic tagging method. To separate the signal $B\to X_{u} \ell \nu$ decay from the background events which are dominated by $B\to X_{c} \ell \nu$, an event classification with boosted decision trees (BDTs) is utilized. After applying the selection of the BDT classifier, a binned likelihood fit is performed to extract the signal yield, where the systematic uncertainties are incorporated via nuisance-parameter constraints. This analysis applied five separate fits to measure the partial branching fractions of three kinematic regions. The $|V_{ub}|$ is extracted based on the measured partial branching fractions with the average $B$ meson lifetime and the state-of-the-art theory predictions on decay rates. The arithmetic average of the most precise determinations for the phase-space region $E^{B}_{\ell}>1$ GeV is $\left|V_{u b}\right|=(4.10 \pm 0.09_{\texttt{stat}} \pm 0.22_{\texttt{syst}} \pm 0.15_{\texttt{theo}}) \times 10^{-3}$. This result is compatible with world averages of exclusive results $\left|V_{u b}^{\text {excl.}}\right|=(3.67 \pm 0.09 \pm 0.12) \times 10^{-3}$ \cite{Amhis:2019ckw} within 1.3 standard deviations and with the value expected from CKM unitarity from a global fit of Ref.~\cite{Charles:2004jd} of $\left|V_{u b}\right|=(3.62^{+0.11}_{-0.08}) \times 10^{-3}$ within 1.6 standard deviations.

Using a similar reconstruction strategy, we also reported the first measurements of differential spectra of inclusive $B \to X_{u} \ell \nu$ decays \cite{cao:2021prl} as a function of the lepton energy $E^{B}_{\ell}$ in the $B$ rest frame, the four-momentum-transfer squared $q^{2}$, light-cone momenta $P_\pm = \left( E^{B}_X \mp |\mathbf{p}^{B}_X|  \right)$, the hadronic mass $M_{X}$, and the hadronic mass squared $M_{X}^{2}$. These measurements pave the way for future direct determinations of the shape function and $|V_{ub}|$, as proposed by Refs.~\cite{Bernlochner:2020jlt,Gambino:2016fdy}. These novel analyses will offer insights into the persistent tensions regarding the value of $|V_{ub}|$ from inclusive and exclusive determinations.

\section{Combined measurements on $|V_{xb}|$}

\subsection{Ratio of exclusive and inclusive $|V_{ub}|$}

The first simultaneous determination of $|V_{ub}|$ using inclusive and exclusive decays is performed at the Belle experiment~\cite{jointVub}. The event reconstruction strategies are inherited from the partial branching fraction measurement~\cite{cao:2021prd}. To disentangle exclusive $B \to \pi  \ell \nu$ decays from other inclusive $B \to X_u  \ell, \nu$ events and backgrounds, we employ a two-dimensional fit. This fitting approach takes into account the number of charged pions in the hadronic $X_u$ system and the four-momentum transfer $q^2$ between the $B$ and $X_u$ system. We constrain the BCL expansion coefficients of $B \to \pi \, \ell \nu$ form factors to the LQCD values of Ref.~\cite{FLAG:2021npn}, combining LQCD calculations from several groups~\cite{FermilabLattice:2015mwy,Flynn:2015mha}. The additive and multiplicative systematic uncertainties are considered in the likelihood fit by adding bin-wise nuisance parameters for each template. The parameters are constrained to a multinormal Gaussian distribution with a covariance reflecting the sum of all considered systematic effects, and the correlation structure between templates originating from shared sources is taken into account. In the ratio of exclusive and inclusive $|V_{ub}|$ values many systematic uncertainties such as the tagging calibration or the lepton identification uncertainties cancel and one can directly test the SM expectation of unity. 

Our findings indicate ratios that align closely with this expectation but exhibit a deviation of 1.5 standard deviations from the ratio of the current world averages of inclusive and exclusive $|V_{ub}|$. This tension decreases to 1.2 standard deviations when including the constraint based on the full theoretical and experimental knowledge of the $B \to \pi  \ell \nu$ form factor shape. With this setup, we obtain the results on $\left|V_{ub}^{\mathrm{excl.}} \right| = (3.78 \pm 0.23 \pm 0.16 \pm 0.14)\times 10^{-3}$ and $\left|V_{ub}^{\mathrm{incl.}} \right| = (3.88 \pm 0.20 \pm 0.31 \pm 0.09)\times 10^{-3}$, with the uncertainties being the statistical, systematic, and theoretical errors. The ratio of $\left|V_{ub}^{\mathrm{excl.}} \right| / \left|V_{ub}^{\mathrm{incl.}} \right| = 0.97 \pm 0.12$ is compatible with unity. Moreover, the averaged $|V_{ub}|$ derived from our inclusive and exclusive values, using LQCD and additional experimental information, is $|V_{ub}|  =  (3.84 \pm 0.26)\times 10^{-3}$. This result is compatible with the expectation from CKM unitarity of Ref.~\cite{CKMfitter2021} of $|V_{ub}^{\mathrm{CKM}}| = (3.64 \pm 0.07)\times 10^{-3}$ within 0.8 standard deviations.

%% Marcel's 
\subsection{Ratio of partial branching fractions of $B \to X_u  \ell \nu$ and $B \to X_c  \ell \nu$}

The semileptonic inclusive decays $B \to X_u  \ell \nu$ and $B \to X_c  \ell  \nu$ are analyzed using the full Belle sample, comprising $772 \times 10^6$ $B$ meson pairs collected at the $\Upsilon(4S)$ resonance. The hadronic tagging algorithm \cite{Keck:2018lcd} developed for Belle~II is employed. The $B \to X_c  \ell \nu$ modeling is modified using sideband data categorized by the Kaon multiplicity. The $B \to X_u  \ell  \nu$ yields are extracted in a two-dimensional fit on the squared four-momentum transfer $q^{2}$ and the charged lepton energy in the B meson rest frame $p_{\ell}^{B}$, and the $B \to X_c  \ell  \nu$ yields are obtained by subtracting other contributions in the total $B \to X \ell \nu$ sample. The study focused on the partial phase space region where $p_{\ell}^{B} > 1 \, \mathrm{GeV}$, encompassing fractions of $\epsilon^{u}_{\Delta} = 86\%$ and $\epsilon^{c}_{\Delta} = 79\%$ of the $B \to X_u  \ell \nu$ and $B \to X_c  \ell \nu$ decays, respectively. Our preliminary result for the partial branching faction ratio is $\Delta\mathcal{B}(B \to X_{u}\ell\nu)/\Delta\mathcal{B}(B \to X_{c}\ell\nu) = 1.95(1 \pm 8.4\%_{\mathrm{stat}} \pm 7.8\%_{\mathrm{syst}})\times 10^{-2}$. This ratio provides insight into the inclusive $|V_{ub}|/|V_{cb}|$ ratio, with theoretical inputs of partial decay rates for both decays. Furthermore, by taking the external normalization of $\Delta \mathcal{B}(B \to X_c \ell  \nu)$, we find the resulting $|V_{ub}|$ value is in good agreement with the previous Belle measurement~\cite{cao:2021prd}.

\section{Summary}
The Belle and Belle~II experiments have provided many new results recently and will be helpful in examining the long-standing $|V_{xb}|$ puzzle. Continuous efforts from experiment and theory are still needed, e.g. the discrepancies in LQCD predictions and measured $B \to D^{*} \ell \nu$ form factors need further in-depth investigation. On the other hand, the experimental results of exclusive $|V_{cb}|$ imply a better agreement with the recent inclusive extractions. The new simultaneous determinations of exclusive and inclusive $|V_{ub}|$, and the inclusive $|V_{ub}|/|V_{cb}|$ ratio measurement, offer additional valuable insights into this puzzle but both require a larger dataset to achieve greater precision. Beyond these important results, the accumulated knowledge on MC modeling, analysis techniques, etc. will be beneficial for future measurements by Belle~II or LHCb.

\bibliographystyle{apsrev4-1}
\bibliography{ref}

\end{document}